\documentclass[english,aps,manuscript,prd,showkeys,showpacs]{revtex4}
\usepackage{lmodern}

\usepackage[]{fontenc}
\usepackage[latin9]{inputenc}
\usepackage{amsmath}
\usepackage{amssymb}

\makeatletter
\@ifundefined{textcolor}{}
{%
 \definecolor{BLACK}{gray}{0}
 \definecolor{WHITE}{gray}{1}
 \definecolor{RED}{rgb}{1,0,0}
 \definecolor{GREEN}{rgb}{0,1,0}
 \definecolor{BLUE}{rgb}{0,0,1}
 \definecolor{CYAN}{cmyk}{1,0,0,0}
 \definecolor{MAGENTA}{cmyk}{0,1,0,0}
 \definecolor{YELLOW}{cmyk}{0,0,1,0}
 }

\makeatother

\makeatother

\usepackage{babel}

\begin{document}

\title{Hidden Conformal Symmetry of Rotating Black Hole with four Charges}

\author{Kai-Nan Shao}

\email{shaokn@gmail.com}

\affiliation{Zhejiang Institute of Modern Physics, Zhejiang University, Hangzhou,
310027, China}

\author{Zhibai Zhang}

\email{zhibai.zhang1987@gmail.com}

\thanks{corresponding author.}

\affiliation{Kavli Institute for Theoretical Physics China, CAS, Beijing 100190,
China}
\begin{abstract}
Kerr/CFT correspondence exhibits many remarkable connections between
the near-horizon Kerr black hole and a conformal field theory (CFT).
Recently, Castro, Maloney and Strominger showed that a hidden conformal
symmetry exists in the solution space of a Kerr black hole. In this
paper we investigate a rotating black hole with four independent $U\left(1\right)$
charges derived from string theory which is known as the four-dimensional
Cvetic-Youm solution, and we prove that the same hidden conformal
symmetry also holds. We obtain the exact blackhole entropy through
the temperatures derived. The entropy and absorption cross section
agree with the previous results {[}M. Cvetic and F. Larsen, Nucl.
Phys. B506, 107 (1997).{]} and {[}M. Cvetic and F. Larsen, J. High
Energy Phys. 09 (2009) 088.{]}. In addition, we clarify a previous
explanation on the temperatures of the Cvetic-Youm solution's dual
CFT. This work provides more robust derivation of the hidden conformal
symmetry as well as Kerr/CFT correspondence. 
\end{abstract}



\maketitle

\section{Introduction}

In \cite{KerrCFT0809.4266}, a duality between a rotating black hole
and conformal field theory, known as Kerr/CFT correspondence was shown
by Guica, Hartman, Song, and Strominger. The derivation of Kerr/CFT
correspondence depends highly on the parameters describing the black
hole. To obtain this correspondence there should be at least one more
parameter other than the black hole mass $M$. Lots of evidences show
that black holes parametrized by more than one variable share this
duality. The extension from a Kerr black hole to black holes parametrized
by more than angular momentum $J$ corresponds to the ideas in \cite{fiveproblem}.
For more works on Kerr/CFT correspondence and its generalizations,
see \cite{KerrCFT} .

Recently, a hidden conformal symmetry of a Kerr black hole was found
by Castro, Maloney, and Strominger \cite{CMS2010a}. Through inspection
on the equation of motion of a test particle propagating in the Kerr
black hole background, a $SL(2,\mathbb{R})_{L}\times SL(2,\mathbb{R})_{R}$
symmetry was found in the radial part of the solution. Although this
symmetry is broken by angular coordinate identification, the correct
entropy and absorption cross section can still be given by the temperatures
$T_{L}$ and $T_{R}$, which implies Kerr/CFT correspondence. And
it shows that the hidden conformal symmetry, although not globally
defined, has provided enough information supporting the duality.

After this interesting evidence was found, several works considering
different types of Kerr-like black holes have been done\cite{hidden,BChen,hidden2,BChenAdS,hidden3}.
Without any violation, the hidden conformal symmetry exists in all
of these black holes, and the conformal field theory (CFT) interpretations
all give the right results. In this paper, we consider the four-dimensional
Cvetic-Youm solution, a four-dimensional rotating black hole with
four independent charges derived from toroidally compactified string
theory\cite{Cvetic:1996kv}. The details of this solution were given
in\cite{Greybody,GreybodyKerrCFT}. We show that the hidden conformal
symmetry still holds by the solution space of black hole of this type,
and the CFT description is correct. A comparison between the temperatures
we derived and the previous ones is also given. From that we make
a clarification about an interpretation in\cite{Greybody,GreybodyKerrCFT}

This paper is organized as follows. In Sec. \ref{sec:Massless-scalar-wave}
we give a description of the parametrization of a rotating black hole
with four charges and the wave equation of a particle propagating
in the black hole background. In Sec. \ref{sec:The--symmetry} we
prove the existence of $SL(2,\mathbb{R})_{L}\times SL(2,\mathbb{R})_{R}$
symmetry in the wave equation. In Sec. \ref{sec:CFT-interpretation-of}
we give the CFT interpretation. Section \ref{sec:Discussion} is for
general discussions.

\section{Massless scalar wave equation in the black hole background\label{sec:Massless-scalar-wave}}

The Cvetic-Youm solution is a rotating black hole with four charges
derived from toroidally compactified string theory \cite{Cvetic:1996kv,Greybody,GreybodyKerrCFT}.
It is parametrized by the mass $M$, angular momentum $J$, and four
independent charges $Q_{i}$,$i=1,2,3,4$. We adopt the convention
in \cite{Greybody,GreybodyKerrCFT} and use $\mu$, $\delta_{i}$,
and $l$ as the new parametrization. From this the old parameters
are given, respectively, as

\begin{equation}
8G_{4}M=\frac{1}{2}\mu\sum_{i=1}^{4}\cosh2\delta_{i}\quad,\end{equation}

\begin{equation}
8G_{4}Q_{i}=\frac{1}{2}\mu\sinh2\delta_{i}\quad,\qquad i=1,2,3,4\quad,\end{equation}

\begin{equation}
8G_{4}J=\frac{1}{2}\mu l\left(\prod_{i=1}^{4}\cosh\delta_{i}-\prod_{i=1}^{4}\sinh\delta_{i}\right)\quad.\end{equation}
 The gravitational coupling constant in four dimensions is $G_{4}=\frac{1}{8}$
which corresponds to $(2\pi)^{6}(\alpha^{\prime})^{4}/V_{6}$ in string
units.

To check the conformal symmetry in the solution space, we set up a
neutral, massless scalar particle propagating in the black hole background.
The solution space is given by the Klein-Gordon equation\begin{equation}
\frac{1}{\sqrt{-g}}\partial_{\mu}\left(\sqrt{-g}g^{\mu\nu}\partial_{\nu}\Phi\right)=0\quad.\end{equation}
 where the metric was given in \cite{Cvetic:1996kv}. This equation
is more conveniently expressed in term of the dimensionless radial
coordinate\begin{equation}
x\equiv\frac{r-\frac{1}{2}(r_{+}+r_{-})}{r_{+}-r_{-}}\quad.\end{equation}
 Here $r$ is the radial coordinate, $r_{+}$ and $r_{-}$ are radii
of the outer and inner horizons. The horizons can be written as $x=\pm\frac{1}{2}$,
respectively, and the asymptotic space corresponds to the large $x$
region. In terms of $\mu,\, l,\,\delta_{i}$, the surface accelerations
of two horizons $\kappa_{\pm}$ and the angular velocity $\Omega$
are \begin{equation}
\frac{1}{\kappa_{\pm}}=\frac{\mu^{2}}{2\sqrt{\mu^{2}-l^{2}}}\left(\prod_{i}\cosh\delta_{i}+\prod_{i}\sinh\delta_{i}\right)\pm\frac{1}{2}\mu\left(\prod_{i}\cosh\delta_{i}-\prod_{i}\sinh\delta_{i}\right)\quad,\label{eq:kapa}\end{equation}
 \begin{equation}
\frac{\Omega}{\kappa_{+}}=\frac{l}{\sqrt{\mu^{2}-l^{2}}}\quad.\label{eq:kapaome}\end{equation}
 and in spherical coordinates the wave function reads\begin{equation}
\Phi\equiv\Phi_{r}\left(x\right)\chi\left(\theta\right)e^{-i\omega t+im\phi}=\Phi\left(x,\theta\right)e^{-i\omega t+im\phi}\quad.\end{equation}
 The wave equation can be written as \cite{Greybody,GreybodyKerrCFT}
\begin{eqnarray}
\frac{\partial}{\partial x}(x^{2}-\frac{1}{4})\frac{\partial}{\partial x}\Phi\left(x,\theta\right)+\frac{1}{4}\biggl[x\Delta^{2}\omega^{2}+xM\Delta\omega^{2}-4\tilde{\Lambda}\nonumber \\
\frac{1}{x-\frac{1}{2}}\left(\frac{\omega}{\kappa_{+}}-m\frac{\Omega}{\kappa_{+}}\right)^{2}-\frac{1}{x+\frac{1}{2}}\left(\frac{\omega}{\kappa_{-}}-m\frac{\Omega}{\kappa_{+}}\right)^{2}\biggl]\Phi\left(x,\theta\right) & = & 0\quad,\label{eq:fullequa}\end{eqnarray}
 where $\tilde{\Lambda}$ is the angular part of the equation\begin{equation}
\tilde{\Lambda}=-\frac{1}{\sin\theta}\frac{\partial}{\partial\theta}\sin\theta\frac{\partial}{\partial\theta}+\frac{m^{2}}{\sin^{2}\theta}-\frac{1}{16}l^{2}\omega^{2}\cos^{2}\theta-\frac{1}{16}\mu^{2}\omega^{2}\left(1+\sum_{i<j}\cosh2\delta_{i}\cosh2\delta_{j}\right)\quad.\end{equation}
 Note that by taking the four charges to zero, which is realized by
taking $\delta_{i}=0$, the angular equation reduces to the form in
Kerr's case.

To investigate the equation, we take the near-region limit introduced
in \cite{CMS2010a}. The near region is where the conformal structure
appears. When the wavelength of the test particle is large enough
compared to the radius curvature\begin{equation}
\omega M\ll1\quad,\end{equation}
 the near region is \begin{equation}
r\ll\frac{1}{\omega}\quad.\label{eq:limit}\end{equation}
 Note that the {}``near'' should not be confused with the one in
{}``near horizon''. Actually, the near region could be arbitrarily
large, as discussed in \cite{CMS2010a}. In this near-region limit
we can neglect the higher order $\omega^{2}$ terms in (\ref{eq:fullequa}),
and therefore the angular part of the equation reduces to the Laplacian
in spherical coordinates \[
\left[\frac{1}{\sin\theta}\partial_{\theta}\left(\sin\theta\partial_{\theta}\right)-\frac{m^{2}}{\sin^{2}\theta}\right]\chi\left(\theta\right)=-\Lambda\chi\left(\theta\right)\quad.\]
 with the constant \[
\Lambda=\tilde{j}\left(\tilde{j}+1\right)\quad.\]
 The radial part of Eq. (\ref{eq:fullequa}) becomes\begin{equation}
\frac{\partial}{\partial x}(x^{2}-\frac{1}{4})\frac{\partial}{\partial x}\Phi_{r}+\frac{1}{4}\biggl[\frac{1}{x-\frac{1}{2}}\left(\frac{\omega}{\kappa_{+}}-m\frac{\Omega}{\kappa_{+}}\right)^{2}-\frac{1}{x+\frac{1}{2}}\left(\frac{\omega}{\kappa_{-}}-m\frac{\Omega}{\kappa_{+}}\right)^{2}\biggl]\Phi_{r}=\tilde{j}(\tilde{j}+1)\Phi_{r}\quad.\label{eq:equa}\end{equation}
 One should notice that, after taking the near-region limit, although
the expressions do not contain $\delta_{i}$ explicitly, the variables
$\omega,$ $\Omega,$ $\kappa_{\pm}$, and $r_{\pm}$ above are determined
by parameters $\mu,$ $l,$ and the four charge parameters$\delta_{i}$.
At this stage, the form of the equation is similar to the one in Kerr's
case. By this observation, we use the formula in \cite{CMS2010a}
to check the operator.

\section{The $SL(2,\mathbb{R})_{L}\times SL(2,\mathbb{R})_{R}$ symmetry in
the solution space\label{sec:The--symmetry}}

Following Ref.\cite{CMS2010a}, we reconstruct the operator in the
above wave equation, by showing that this operator could be viewed
as the Casimir of a $SL\left(2,\mathbb{R}\right)$ group. First, one
should employ the conformal coordinates\begin{eqnarray}
\omega^{+} & = & \sqrt{\frac{2x-1}{2x+1}}e^{2\pi T_{R}\phi}\quad,\nonumber \\
\omega^{-} & = & \sqrt{\frac{2x-1}{2x+1}}e^{2\lambda_{L}t+2\pi T_{L}\phi}\quad,\nonumber \\
y & = & \sqrt{\frac{2}{2x+1}}e^{\lambda_{L}t+\pi(T_{L}+T_{R})\phi}\quad,\end{eqnarray}
 where\begin{equation}
T_{R}=\frac{\kappa_{+}}{2\pi\Omega}\quad,\label{eq:rtem}\end{equation}
 \begin{equation}
T_{L}=\frac{\kappa_{+}(\kappa_{-}+\kappa_{+})}{2\pi\Omega(\kappa_{-}-\kappa_{+})}\quad,\label{eq:ltem}\end{equation}
 and \begin{equation}
\lambda_{L}=\frac{\kappa_{+}\kappa_{-}}{\kappa_{-}-\kappa_{+}}\quad.\end{equation}
 These three variables contain the $\delta_{i}$-dependence implicitly,
and are different from the ones in Kerr black hole. $T_{L}$ and $T_{R}$
are the temperatures of the dual CFT. Defining the left and right
temperatures is the key point in the whole formula. The right decision
which presents the hidden conformal symmetry will also match the CFT
description, and vice versa. This is an important feature of the hidden
conformal symmetry.

Then one can define the local vectors\begin{eqnarray}
H_{1} & = & i\partial_{+}\quad,\nonumber \\
H_{0} & = & i\left(\omega^{+}\partial_{+}+\frac{1}{2}y\partial_{y}\right)\quad,\nonumber \\
H_{-1} & = & i\left(\omega^{+2}\partial_{+}+\omega^{+}y\partial_{y}-y^{2}\partial_{-}\right)\quad,\end{eqnarray}
 and \begin{eqnarray}
\bar{H}_{1} & = & i\partial_{-}\quad,\nonumber \\
\bar{H}_{0} & = & i\left(\omega^{-}\partial_{-}+\frac{1}{2}y\partial_{y}\right)\quad,\nonumber \\
\bar{H}_{-1} & = & i\left(\omega^{-2}\partial_{-}+\omega^{-}y\partial_{y}-y^{2}\partial_{+}\right)\quad.\end{eqnarray}
 Clearly, they satisfy two sets of $SL(2,\mathbb{R})$ Lie algebra
respectively\[
\left[H_{0,}H_{\pm1}\right]=\mp iH_{\pm1}\quad,\qquad\left[H_{-1,}H_{1}\right]=-2iH_{0}\quad,\]
 \begin{equation}
\left[\bar{H}_{0,}\bar{H}_{\pm1}\right]=\mp i\bar{H}_{\pm1}\quad,\qquad\left[\bar{H}_{-1,}\bar{H}_{1}\right]=-2i\bar{H}_{0}\quad.\end{equation}
 The quadratic Casimir of this algebra is\begin{eqnarray}
H^{2} & = & \bar{H}^{2}=-H_{0}^{2}+\frac{1}{2}\left(H_{1}H_{-1}+H_{-1}H_{1}\right)\nonumber \\
 & = & \frac{1}{4}\left(y^{2}\partial_{y}^{2}-y\partial_{y}\right)+y^{2}\partial_{+}\partial_{-}\label{eq:casi}\end{eqnarray}
 which we identify as the operator that appears in the left-hand side
of Eq.(\ref{eq:equa}). To see this, one needs to rewrite the variables
in terms of the $(t,x,\phi)$ coordinates. In our case the vector
fields are\begin{eqnarray}
H_{1} & = & \frac{ie^{-2\pi T_{R}\phi}\sqrt{4x^{2}-1}}{2}\partial_{x}+\frac{ie^{-2\pi T_{R}\phi}x}{\pi T_{R}\sqrt{4x^{2}-1}}\partial_{\phi}-\frac{ie^{-2\pi T_{R}\phi}(T_{R}+2T_{L}x)}{2T_{R}\text{ }\sqrt{4x^{2}-1}\lambda_{L}}\partial_{t}\quad,\nonumber \\
H_{0} & = & \frac{i\left(\lambda_{L}\partial_{\phi}-\pi T_{L}\partial_{t}\right)}{2\pi T_{R}\lambda_{L}}\quad,\nonumber \\
H_{-1} & = & -\frac{ie^{2\pi T_{R}\phi}\sqrt{4x^{2}-1}}{2}\partial_{x}+\frac{ie^{2\pi T_{R}\phi}x}{\pi T_{R}\sqrt{4x^{2}-1}}\partial_{\phi}-\frac{ie^{2\pi T_{R}\phi}(T_{R}+2T_{L}x)}{2T_{R}\sqrt{4x^{2}-1}\lambda_{L}}\partial_{t}\quad,\end{eqnarray}
 and the antiholomorphic part\emph{ }is\begin{eqnarray}
\bar{H}_{1} & = & \frac{ie^{-2\left(\pi T_{L}\phi+\lambda_{L}t\right)}\sqrt{4x^{2}-1}}{2}\partial_{x}-\frac{ie^{-2\left(\pi T_{L}\phi+\lambda_{L}t\right)}}{2\pi\sqrt{4x^{2}-1}T_{R}}\partial_{\phi}+\frac{ie^{-2\left(\pi T_{L}\phi+\lambda_{L}t\right)}\left(T_{L}+2xT_{R}\right)}{2\sqrt{4x^{2}-1}T_{R}\lambda_{L}}\partial_{t}\quad,\nonumber \\
\bar{H}_{0} & = & \frac{i}{2\lambda_{L}}\partial_{t}\quad,\nonumber \\
\bar{H}_{-1} & = & -\frac{ie^{2\pi T_{L}\phi+2\lambda_{L}t}\sqrt{4x^{2}-1}}{2}\partial_{x}-\frac{ie^{2\pi T_{L}\phi+2\lambda_{L}t}}{2\pi\sqrt{4x^{2}-1}T_{R}}\partial_{\phi}+\frac{ie^{2\pi T_{L}\phi+2\lambda_{L}t}\left(T_{L}+2xT_{R}\right)}{2\sqrt{4x^{2}-1}T_{R}\lambda_{L}}\partial_{t}\quad.\end{eqnarray}
 Thus here $H^{2}$ on the left-hand side of the wave equation is
exactly the $SL(2,\mathbb{R})$ \begin{eqnarray}
H^{2} & = & -\frac{1}{4\pi^{2}T_{R}^{2}\left(4x^{2}-1\right)}\partial_{\phi}^{2}-\frac{T_{L}^{2}+T_{R}^{2}+4T_{L}T_{R}x}{4T_{R}^{2}\left(4x^{2}-1\right)\lambda_{L}^{2}}\partial_{t}^{2}\nonumber \\
 &  & +\frac{T_{L}+2T_{R}x}{2\pi T_{R}^{2}\lambda_{L}\left(4x^{2}-1\right)}\partial_{t}\partial_{\phi}+\left(x^{2}-\frac{1}{4}\right)\partial_{x}^{2}+2x\partial_{x}\quad.\end{eqnarray}
 The original scalar field wave equation (\ref{eq:equa}) can be rewritten
as\begin{equation}
H^{2}\Phi=\bar{H}^{2}\Phi=\tilde{j}\left(\tilde{j}+1\right)\Phi\quad.\end{equation}
 By writing the wave equation we see that the hidden conformal symmetry
is now manifest. The vectors $\left\{ H_{0,\pm1}\right\} _{L}$ and
$\left\{ \bar{H}_{0,\pm1}\right\} _{R}$ together generate a $SL\left(2,\mathbb{R}\right)_{L}\times SL\left(2,\mathbb{R}\right)_{R}$
algebra. The weights of the fields are then\begin{equation}
h_{L}=\tilde{j}\quad,\qquad h_{R}=\tilde{j}\quad.\end{equation}

\section{CFT interpretation of $T_{L}$ and $T_{R}$\label{sec:CFT-interpretation-of}}

\subsection{Temperature and Entropy}

Having proven that, after taking the near region limit (\ref{eq:limit}),
the operator in the radial part of wave equation can be identified
as the Casimir of a $SL(2,\mathbb{R})_{L}\times SL(2,\mathbb{R})_{R}$
algebra. A reasonable next step is to consider that the near region
of the black hole should be dual to a $\left(T_{L},T_{R}\right)$
finite temperature state in a CFT. As an examination, we calculate
the black hole's entropy microscopically using the Cardy formula,
\begin{equation}
S=\frac{\pi^{2}}{3}\left(c_{L}T_{L}+c_{R}T_{R}\right)\quad.\end{equation}
 However, the $SL(2,\mathbb{R})_{L}\times SL(2,\mathbb{R})_{R}$ is
a Virasoro algebra without central charge. This situation is the same
as the case of Kerr black hole. In the original Kerr/CFT correspondence,
the central charges are derived in the near horizon region of extremal
Kerr black hole \cite{KerrCFT0809.4266}. In the same sense as in
\cite{CMS2010a}, we conjecture that, the central charges will keep
valid both from an extremal black hole to a nonextremal one, and from
near horizon to the near region defined above. Based on that, we will
use the central charges derived previously and temperatures obtained
in this paper. For extreme Kerr, from the asymptotic symmetry group
the central charges are given as\begin{equation}
c_{R}=c_{R}=12J\quad.\end{equation}
 Cooperating with the temperatures given in (\ref{eq:rtem}) and (\ref{eq:ltem}),
from the Cardy formula the entropy of the black hole is given as\begin{eqnarray}
S & = & \frac{\pi^{2}}{3}\left(c_{L}T_{L}+c_{R}T_{R}\right)\nonumber \\
 & = & 4\pi J\left(\frac{\kappa_{+}}{\Omega}\right)\left(\frac{\kappa_{-}}{\kappa_{-}-\kappa_{+}}\right)\quad.\end{eqnarray}
 Using (\ref{eq:kapa}) and (\ref{eq:kapaome}) we have\begin{equation}
S=\frac{2\pi}{8G_{4}}\left[\frac{1}{2}\mu^{2}\left(\prod_{i}\cosh\delta_{i}+\prod_{i}\sinh\delta_{i}\right)+\frac{1}{2}\mu\sqrt{\mu^{2}-l^{2}}\left(\prod_{i}\cosh\delta_{i}-\prod_{i}\sinh\delta_{i}\right)\right]\quad.\end{equation}
 According to \cite{Greybody,GreybodyKerrCFT}, the Bekenstein-Hawking
entropy of Cvetic-Youm solution is\begin{eqnarray}
S & = & \frac{A}{4G_{N}}\nonumber \\
 & = & 2\pi\left[\frac{1}{2}\mu^{2}\left(\prod_{i}\cosh\delta_{i}+\prod_{i}\sinh\delta_{i}\right)+\frac{1}{2}\mu\sqrt{\mu^{2}-l^{2}}\left(\prod_{i}\cosh\delta_{i}-\prod_{i}\sinh\delta_{i}\right)\right]\end{eqnarray}
 Apparently, the two entropies are equal to each other after setting
$G_{4}=\frac{1}{8}$. Thus as we promised, $T_{L}$ and $T_{R}$ indeed
give the correct entropy, and the near region of the black hole should
be dual to a CFT.

Additionally, we would like to point out that there is an interesting
comparison between the $L-$ and $R-$ temperature defined in \cite{Greybody,GreybodyKerrCFT}
and the $T_{L}$ and $T_{R}$ we used in this paper. These two pairs
of temperatures differ by a factor $\mathcal{R}_{4}$. $T_{L}$ and
$T_{R}$ in this paper emerge as a direct result of the hidden conformal
symmetry, and they match the Cardy formula perfectly. In \cite{GreybodyKerrCFT},
to get the correct entropy microscopically, the factor \[
\mathcal{R}_{4}=\mu\left(\prod_{i=1}^{4}\cosh\delta_{i}+\prod_{i=1}^{4}\sinh\delta_{i}\right)\]
 is embedded into the Cardy formula. And they considered the result
derived in this way as a phenomenological model describing the black
hole. Based on the result we interpret it in another way: by putting
this factor $\mathcal{R}_{4}$ into the $L$ and $R$ temperature
instead, they become the exact $T_{L}$ and $T_{R}$ above in this
paper, and then the Cardy formula will make sense without any modification.
Thus, if the Cardy formula gives the correct entropy, then the temperatures
will derive the hidden conformal symmetry. Therefore, this is an important
evidence of the hidden conformal symmetry.

\subsection{Absorption Probabilities}

Next we analyze the black hole's absorption probabilities for a massless
neutral scalar in the near region. A systematic way to exam the CFT
interpretation is to check the behavior of the test particle's solution\cite{CMS2010a}.
In our case, it is

\begin{eqnarray}
\Phi_{0}^{in} & = & \left(\frac{x-\frac{1}{2}}{x+\frac{1}{2}}\right)^{-\frac{i\beta_{H}\left(\omega-m\Omega\right)}{4\pi}}\left(x+\frac{1}{2}\right)^{-1-\tilde{j}}\nonumber \\
 &  & F\left(1+\tilde{j}-i\frac{\beta_{R}\omega-2\beta_{H}m\Omega}{4\pi},1+\tilde{j}-i\frac{\beta_{L}\omega}{4\pi},1-i\frac{\beta_{H}}{4\pi}\left(\omega-m\Omega\right),\frac{x-\frac{1}{2}}{x-\frac{1}{2}}\right)\quad.\end{eqnarray}
 Here the symbols above are defined as\begin{equation}
\beta_{H}=\frac{2\pi}{\kappa_{+}}\quad,\qquad\beta_{R}=\frac{2\pi}{\kappa_{+}}+\frac{2\pi}{\kappa_{-}}\quad,\qquad\beta_{L}=\frac{2\pi}{\kappa_{+}}-\frac{2\pi}{\kappa_{-}}\quad.\end{equation}
 From \cite{Greybody,GreybodyKerrCFT}, the asymptotic form of $\Phi$
is\begin{eqnarray}
\Phi_{0}^{in} & \sim & Ax^{\tilde{j}}+Bx^{-1-\tilde{j}}\nonumber \\
 & \sim & x^{-1-\tilde{j}}\frac{\Gamma\left(1-i\frac{\beta_{H}\left(\omega-m\Omega\right)}{2\pi}\right)\Gamma\left(-1-2\tilde{j}\right)}{\Gamma\left(\tilde{j}-i\frac{\beta_{L}\omega}{2\pi}\right)\Gamma\left(\tilde{j}-i\frac{\beta_{R}\omega-2\beta_{H}m\Omega}{2\pi}\right)}+x^{\tilde{j}}\frac{\Gamma\left(1-i\frac{\beta_{H}\left(\omega-m\Omega\right)}{2\pi}\right)\Gamma\left(1+2\tilde{j}\right)}{\Gamma\left(1+\tilde{j}-i\frac{\beta_{L}\omega}{2\pi}\right)\Gamma\left(1+\tilde{j}-i\frac{\beta_{R}\omega-2\beta_{H}m\Omega}{2\pi}\right)}\quad.\end{eqnarray}
 At the outer boundary of the matching region\begin{equation}
\Phi_{0}^{in}\sim x^{\tilde{j}}\frac{\Gamma\left(1-i\frac{\beta_{H}\left(\omega-m\Omega\right)}{2\pi}\right)\Gamma\left(1+2\tilde{j}\right)}{\Gamma\left(1+\tilde{j}-i\frac{\beta_{L}\omega}{2\pi}\right)\Gamma\left(1+\tilde{j}-i\frac{\beta_{R}\omega-2\beta_{H}m\Omega}{2\pi}\right)}\quad,\end{equation}
 and thus, from properties of $\Gamma$ function we obtain\begin{eqnarray}
P_{abs} & \sim & \left|A\right|^{-2}\nonumber \\
 & \sim & \sinh\left(\frac{\beta_{H}\left(\omega-m\Omega\right)}{2\pi}\right)\left|\Gamma\left(1+\tilde{j}-i\frac{\beta_{L}\omega}{2\pi}\right)\right|^{2}\left|\Gamma\left(1+\tilde{j}-i\frac{\beta_{R}\omega-2\beta_{H}m\Omega}{2\pi}\right)\right|^{2}\quad.\label{eq:pabs}\end{eqnarray}
 From the first law of thermodynamics\begin{equation}
T_{H}\delta S=\delta M-\Omega\delta J\quad,\end{equation}
 we wish to find the conjugate charges $\delta E_{R}$ and $\delta E_{L}$
such that\begin{equation}
\delta S=\frac{\delta E_{L}}{T_{L}}+\frac{\delta E_{R}}{T_{R}}\quad.\end{equation}
 With the identification chosen as\[
\delta M=\omega\quad,\qquad\delta J=m\quad,\]
 the solution is\[
\delta E_{L}=\frac{\kappa_{-}+\kappa_{+}}{2\Omega\kappa_{-}}\delta M\quad,\]
 \[
\delta E_{R}=\frac{\kappa_{-}-\kappa_{+}}{2\Omega\kappa_{-}}\delta M-\delta J\quad,\]
 We chose the identification of left and right frequencies as\begin{equation}
\delta E_{L}\equiv\omega_{L}\quad,\qquad\delta E_{R}\equiv\omega_{R}\quad,\qquad\end{equation}
 and get the result \begin{equation}
\omega_{L}=\frac{\omega}{2\Omega}\frac{\kappa_{-}+\kappa_{+}}{\kappa_{-}}\quad,\omega_{R}=\frac{\omega}{2\Omega}\frac{\kappa_{-}-\kappa_{+}}{\kappa_{-}}-m\quad.\end{equation}

Putting these back into (\ref{eq:pabs}), the expression of absorption
probability turns out to be\begin{equation}
P_{abs}\sim T_{L}^{2h_{L}-1}T_{R}^{2h_{R}-1}\sinh\left(\frac{\omega_{L}}{2T_{L}}+\frac{\omega_{R}}{2T_{R}}\right)\left|\Gamma\left(h_{L}+\frac{\omega_{L}}{2\pi T_{L}}\right)\right|^{2}\left|\Gamma\left(h_{R}+\frac{\omega_{R}}{2\pi T_{R}}\right)\right|^{2}\quad.\end{equation}
 which is exactly the form of finite temperature cross section for
CFT.

\section{Discussion\label{sec:Discussion}}

In this paper, we prove the existence of the hidden conformal symmetry
of a rotating black hole with four charges known as the four-dimensional
Cvetic-Youm solution. The duality to a conformal field theory is also
discussed, with the calculations on entropy and absorption cross section
being the same as previous results. The only particle type we consider
in this paper is a scalar, which has a relatively simple wave function.
If the formula used to derive the hidden conformal symmetry is consistent,
it should be also valid in the cases when higher spin particles are
taken into consideration. A calculation on the correlator of photons
and gravitons has been given in \cite{BChen}, while evidences more
than these are expected. The key to show the hidden conformal symmetry
is to construct the operator in the wave function as a Casimir induced
by the conformal coordinates. Wave functions of higher spin particles
in black hole background were studied many years ago \cite{NewmanPenrose}.
These works show that, in general the angular and radial part satisfies
the Teukolsky function,\begin{eqnarray*}
\left[\frac{1}{\sin\theta}\frac{d}{d\theta}\left(\sin\theta\frac{d}{d\theta}\right)+\left(\Lambda_{lm}^{s}-a^{2}\omega^{2}\sin^{2}\theta-2a\omega s\cos\theta-\frac{m^{2}+s^{2}+2ms\cos\theta}{\sin^{2}\theta}\right)\right]\chi^{s}\left(\theta\right) & = & 0\\
\left[\Delta^{-s}\frac{d}{dr}\left(\Delta^{s+1}\frac{d}{dr}\right)+\left(\frac{H^{2}-2is\left(r-M\right)H}{\Delta}+4is\omega r+2am\omega+s\left(s+1\right)-\Lambda_{lm}^{s}\right)\right]\Phi^{s}\left(r\right) & = & 0\end{eqnarray*}
 where $\Delta$ and $H$ are determined by the concrete black hole
configuration, and $s$ is the spin of the test particle. One can
find the condition that $s\neq0$ will make the radial part more complicated
even after neglecting the $\omega^{2}$ terms by the near-region limit.
Modification of the conformal coordinates seems to be a solution,
and we leave this to future works. One would also consider the case
involving the cosmology constant. Discussions on multiple-charge rotating
black hole in anti-de Sitter (AdS) space were given by \cite{Cvetic:1999xp}.
Searching the hidden conformal symmetry in these Kerr-like AdS black
holes' backgrounds could be another interesting topic waiting to be
explored. Results of Kerr-Newman AdS have been given in \cite{BChenAdS}.

On the other hand, as for the near region, it's actually not so \char`\"{}near\char`\"{}.
According to the holographic principle, the black hole's dual field
theory should be constructed on the horizon. Thus it is interesting
to ask what will happen, if we applied more (or different) constraints
to distinguish the regions (which give rise to various limits we can
apply on the wave function), and whether symmetries would appear in
these regions. 
\begin{acknowledgments}
We thank Professor Hong Lü and Dr. Yong-Qiang Wang for help and useful
discussions. We thank Dr. Chih-Hao Fu for modifying the manuscript.
We are grateful to the KITPC, Beijing, for hospitality during this
work. This research is supported in part by the Project of Knowledge
Innovation Program (PKIP) of Chinese Academy of Sciences, Grant No.
KJCX2.YW.W10. Kai-Nan Shao is supported in part by the NNSF of China
Grant No. 90503009, No. 10775116, and 973 Program Grant No. 2005CB724508.\end{acknowledgments}

\end{document}